\documentclass[aps,prb,longbibliography,letterpaper,amsmath,twocolumn]{revtex4-2}

\usepackage{graphicx}
\usepackage{dcolumn}
\usepackage{bm}
\usepackage[colorlinks = true, linkcolor = blue, urlcolor  = blue, citecolor = blue, anchorcolor = blue]{hyperref}
\usepackage[english]{babel}
\usepackage{epsfig}
\usepackage{times}
\begin{document}

\title{Causal optimization method for imaginary-time Green's functions in interacting electron systems}

\author{Mancheon Han}
\email{einnew90@gmail.com}
\author{Hyoung Joon Choi}
\email{h.j.choi@yonsei.ac.kr}
\affiliation{Department of Physics, Yonsei University, Seoul, 03722, Korea}

\begin{abstract}
We develop a causal optimization method that ensures causality in numerical calculations
of Green's functions in interacting electron systems. 
Our method removes noncausality of numerical data by finding causal functions closest to the data.
By testing our method with an exactly calculable model and applying it to practical dynamical mean-field calculations, 
we find that intermediate-frequency behaviors of Green's functions are determined solely by causality,
and noncausal statistical errors are removed very efficiently.
Furthermore, we demonstrate that numerical calculations of the physical branch of the Luttinger-Ward functional
can be stabilized by ensuring causality of the noninteracting Green's function.
Our method and findings provide a basis for improving stability and efficiency of numerical simulations of quantum
many-body systems.
\end{abstract}

\maketitle

\section{Introduction}

Causality is a profound principle that a cause should precede its effects.
In many-body physics, this principle is embodied by the Heaviside step 
function $\theta(t)$ in the definition of the retarded Green's 
function~\cite{Abrikosov,Economou2013}.
For a fermionic Green's function, the step function leads to the Lehmann
representation~\cite{Lehmann} with nonnegative spectral 
function ~\cite{Hettler1998}.
When the Green's function is generalized to a matrix form,
the spectral function is required to be a positive-semidefinite 
matrix~\cite{Hettler1998}.
Moreover, quantities related to the Green's function such as 
the self-energy~\cite{Potthoff2003,Potthoff2007,Staar2014} and the hybridization 
function~\cite{Werner2006:2,Gull2011,Haule2010} should satisfy 
corresponding causality conditions.

Results of practical calculations, however, often contain noncausality.
Such noncausality can be divided into two categories according to its origin.
The first category is from incomplete causality of the theory itself.
For example, extensions of dynamical mean-field theory (DMFT) to consider nonlocal correlations can induce
causality violation~\cite{Okamoto2003,PGJ1994,Avraham1995,Lee2017,Vucicevic2018}.
The second category is from the use of numerical methods that do not enforce the causality.
A representative example of this category is the DMFT~\cite{Metzner1989,Muller1989:a,Muller1989:b,Georges1992,Georges1996} 
with the quantum Monte Carlo (QMC) method~\cite{Blumer2007,Werner2006:1,Rubtsov2005,Gull2011}.
In this case, statistical errors induce noncausality even though DMFT itself is a causal theory~\cite{Hettler1998,Hettler2000}.
In addition, nonstatistical methods for DMFT can induce the causality violation~\cite{Pairault2000,Ago2017}.

There have been many attempts to resolve the noncausality issue
because satisfying causality is crucial for physically consistent calculations~\cite{Haule2010,Okamoto2003}.
Generally, the first category of noncausality can be solved by constructing
causal theories~\cite{Hettler1998,Hettler2000,Kotliar2001,Biroli2004,Backes2020}.
The second category of noncausality can be solved by eliminating 
negative regions
of the spectral function in the case of real-time calculations.
However, in the case of imaginary-time calculations,
no general method has been reported to ensure the causality although
practical finite-temperature many-body calculations are often performed in
imaginary time~\cite{Werner2006:2,Gull2011}.

The Luttinger-Ward functional~\cite{Luttinger1,Baym1961} is
the central building block for many theoretical and numerical approaches 
in correlated electron systems. Recently, this functional is found 
multivalued~\cite{Kozik2015,Rossi2015,Gunnarsson2017,Vucicevic2018,Ajkim2020}, having
unphysical branches of solution. In practical calculations, 
this multivaluedness can make programs converge 
to unphysical solutions~\cite{Kozik2015, Vucicevic2018},
breaking causality~\cite{Vucicevic2018} in
nested cluster DMFT simulations~\cite{Georges1996,Avraham1995,Biroli2004},
for instance.
It was pointed out that constraining the noninteracting Green's function $G_0$ 
to be physical can avoid such unphysical branches
\cite{Kozik2015,Potthoff2003,Stan2015,Eder2014},
but general and practical method is yet to be developed.

In this work, we develop a causal optimization method which ensures causality of imaginary-time and imaginary-frequency Green's functions
and investigate roles of causality on numerical calculations in imaginary time and frequency.
We test our method using an exactly calculable model.
Then, we apply our method to DMFT simulation with
the continuous-time QMC program.
Finally, we show our method suppresses unphysical branches of the Luttinger-Ward functional.

\section{Causal Optimization for fermionic functions}

\subsection{Theoretical framework}

As mentioned in the Introduction, the causality of the fermionic Green's function $G(z)$ 
leads to the Lehmann representation with the nonnegative spectral function
$A(x)$~\cite{Hettler1998}:
\begin{equation}
   G(z) = \int  \frac{A(x)}{z -x} dx,\quad  A(x) \geq 0.  \label{eq:pd}
\end{equation}
In the imaginary time, Eq.~(\ref{eq:pd}) is represented as~\cite{Gubernatis1991}
\begin{equation}
    G(\tau) = -\int  {A(x) n_F(-x) e^{-\tau x}} d x , \quad A(x) \geq 0. \label{eq:tau}
\end{equation}
Here $n_F$ is the Fermi-Dirac distribution function~\cite{Abrikosov},
$0\leq \tau \leq \beta$, and $\beta=(k_BT)^{-1}$. Because $G(\tau)$ is antiperiodic 
in $-\beta \leq \tau \leq \beta$~\cite{Abrikosov}, 
we consider $0 \leq \tau \leq \beta$ only. 
Now, suppose we have some numerical data of $G(\tau)$ obtained from, for example, QMC simulations. 
Then, we consider to find a causal Green's function $G_\textrm{c}(\tau)$
that satisfies Eq.~(\ref{eq:tau}) and minimizes the distance $d$ defined as
\begin{eqnarray}
	\nonumber
    d &=& \frac{1}{\beta}\int_0^{\beta} (G(\tau)-G_\textrm{c}(\tau))^2 d \tau  \\ 
	\label{eq:chi2}
      &=& (k_B T)^2 \sum_{i \omega_n} |G(i \omega_n)-G_\textrm{c}(i\omega_n)|^2.
\end{eqnarray}
Here the imaginary-frequency Green's function $G(i\omega_n)$ is 
at $\omega_n = (2n+1)\pi k_B T$.
If $G(i \omega_n)$ is causal, then $G_\textrm{c}(i \omega_n) = G(i \omega_n)$ trivially.
If not, then $G_\textrm{c}(i \omega_n)$ is the causal function closest to $G(i \omega_n)$.
Once $G_\textrm{c}$ is obtained, we replace $G$ with $G_\textrm{c}$.
This procedure is equivalent to optimizing $G$ 
to the closest causal function, so
we name this method as {\em the causal optimization}.
Since it is not straightforward to use Eq.~(\ref{eq:tau}) during minimization of $d$,
we derive constraints from Eq.~(\ref{eq:tau}).
From Eq.~(\ref{eq:tau}), the Green's function and its derivatives $G^{(n)}(\tau)=d^nG(\tau)/d^n\tau$ satisfy~\cite{BlumerPhD,Vucicevic2018} 
\begin{equation}
     G^{(2k)}(\tau) \leq 0  \text{ for } k = 0, 1, 2, \cdots. \label{eq:ct}
\end{equation}
This is very restrictive constraints in the imaginary time.
We consider to use Eq.~(\ref{eq:ct}) instead of Eq.~(\ref{eq:tau}) during minimization
of $d$. 
Although Eq.~(\ref{eq:ct}) is not a sufficient but a necessary condition of Eq.~(\ref{eq:tau}),
we will present below that this method can be very successful in obtaining the causal function $G_\textrm{c}$
closest to numerical data of $G$, in practice.

Numerical implementation of our causal optimization method needs to consider
$G_\textrm{c}$ and its derivatives
to satisfy Eq.~(\ref{eq:ct}) while minimizing $d$ 
for given values of $G(\tau)$ or $G(i\omega_n)$.
To do this, we employ the cubic-spline 
interpolation~\cite{Boor2,GullPhD,Georges1996,Bergeron2011,Boor1} at the adaptively 
generated grid $\{\tau_i\}$~\cite{Boor1} in $0\leq\tau_i\leq\beta$. This
interpolation yields $G^{(1)}_\textrm{c}(\tau_i)$, $G^{(2)}_\textrm{c}(\tau_i)$, and $G^{(3)}_\textrm{c}(\tau_i)$ as linear combinations of 
$G_\textrm{c}(\tau_i)$. 
Then, we express Eq.~(\ref{eq:ct}) as
\begin{equation}
\begin{array}{ll}
{\displaystyle G_\textrm{c}(\tau_i) \leq0,\quad} 
{\displaystyle G^{(2)}_\textrm{c}(\tau_i) \leq 0,} \\
{\displaystyle G^{(3)}_\textrm{c}(\tau_{i+1})-G^{(3)}_\textrm{c}(\tau_i)\leq 0,}
\end{array}
\label{eq:nct}
\end{equation}
for all $\tau_i$.
Here the third inequality is from nonpositivity of $G_\textrm{c}^{(4)}$.
Because derivatives at $\tau_i$
can be found by linear transforms of $G_\textrm{c}(\tau_i)$, 
Eq.~(\ref{eq:nct}) 
is a set of linear constraints on $G_\textrm{c}(\tau_i)$. 
Meanwhile, $G_\textrm{c}(i\omega_n) = \sum_i K_{ni}G_\textrm{c}(\tau_i)$ with
coefficients $K_{ni}$ determined by the cubic-spline integration~\cite{Press1989,Bergeron2011}. 
Then, 
Eq.~(\ref{eq:chi2}) is 
\begin{equation}
     d = (k_B T)^2 \sum_{i\omega_n}| G(i \omega_n)-
     \sum_i K_{ni} G_\textrm{c}(\tau_i) |^2.  \label{eq:nchi2}
\end{equation}
We employ the
interior-point method for the quadratic programming~\cite{Nocedal,Boyd}
to minimize $d$ 
with respect to $G_\textrm{c}(\tau_i)$ satisfying
Eq.~(\ref{eq:nct}).
Although Eq.~(\ref{eq:nct}) is only up to the fourth derivative, 
it is successful in constraining $G_\textrm{c}$
to be causal during minimization of $d$.
Details of the adaptively generated nonuniform grid, the cubic-spline
interpolation, and the quadratic programming are described in Appendix A.

Our causal optimization method can also be applied to the hybridization function 
$\Delta(z)$ and the self-energy $\Sigma(z)$.
The same condition as Eq.~(\ref{eq:pd})
applies to $\Delta(z)$, while
$\Sigma(z)$ satisfies ~\cite{Potthoff2003,Potthoff2007,Staar2014}
\begin{equation}
    \Sigma(z) = \Sigma_\infty + \int  \frac{A_\Sigma(x)}{z -x} d x,\quad  A_\Sigma(x) \geq 0.  \label{eq:sigpd}
\end{equation}
Here the causal optimization is applicable to the dynamic part,
$\Sigma(i\omega_n)-\Sigma_\infty$.
We can use high-frequency asymptotic coefficients $M_{i}$,  
which satisfy
$G(i \omega_n)\rightarrow{M_0}/{i \omega_n} + {M_1}/{(i \omega_n)^2} +
{M_2}/{(i \omega_n)^3}  + \cdots $ as $\omega_n \rightarrow \infty$, 
as constraints because 
they can be calculated much more accurately than 
imaginary-frequency values~\cite{Haule2007,Gull2011}. 
As for computational efficiency, our method is a light process.
So noncausality of data can be eliminated with 
little additional computation cost.

\begin{figure} 
\epsfxsize=8.5cm
\centering
\centerline{\epsfbox{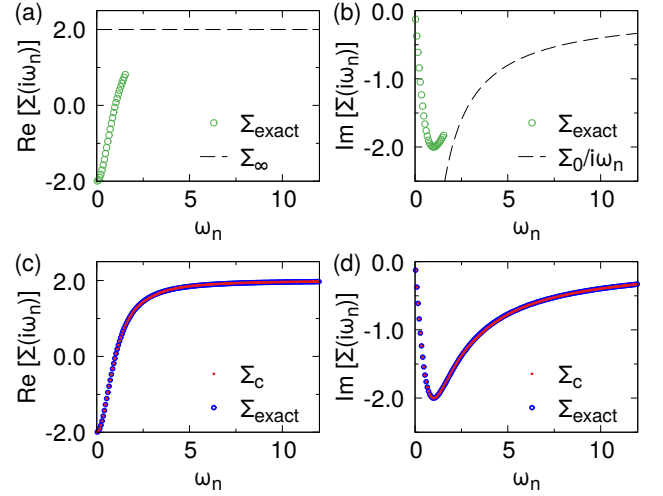}}
\caption {Causal optimization applied to the self-energy given at low frequencies. 
(a) Real and (b) imaginary parts of the exact 
self-energy $\Sigma_\textrm{exact}$
of the Hubbard atom given at low frequencies, shown in green dots.
Dashed lines represent the high-frequency asymptote.
(c) Real and (d) imaginary parts of the 
causal self-energy $\Sigma_{\textrm{c}}$ 
in the entire frequency range, shown in red dots, obtained from our causal 
optimization method using only $\Sigma_\textrm{exact}$
at low frequencies and 
the high-frequency asymptote plotted in (a) and (b). In (c) and (d),
$\Sigma_{\textrm{c}}$ agrees excellently in the entire frequency range
with $\Sigma_{\textrm{exact}}$ plotted in blue dots for comparison.}
\label{fig:1}
\end{figure}

\subsection{Hubbard atom}

To test our method, we first consider the Hubbard atom at temperature $T$
described by Hamiltonian
\begin{equation}
    H = \epsilon (n_\uparrow + n_\downarrow) + U n_\uparrow n_\downarrow. \label{eq:Hhat}
\end{equation}
Here $U$ is the Coulomb interaction strength, $\epsilon$ is the local level,
and $n_\sigma$ is the electron occupation with spin $\sigma$.
The exact Green's function is~\cite{Kajueter1996}
\begin{equation}
   G(i \omega_n) = \frac{n}{i \omega_n -\epsilon - U} + \frac{1-n}{i \omega_n -\epsilon}, \label{eq:hat}
\end{equation}
where $n$ is the number of electrons per spin.
Then, the self-energy is $\Sigma (i \omega_n) = i \omega_n -\epsilon - G^{-1}(i \omega_n)$ from the Dyson equation.
We considered the case of $k_B T=0.01$, $n=0.5$, $\epsilon=-1$, and $U=4$.
To examine the role of causality in determining the self-energy in intermediate imaginary frequency,
we performed the causal optimization of the self-energy using exact self-energy values
at 25 low frequencies and high-frequency asymptotic coefficients $\Sigma_\infty$ and $\Sigma_0$ for
the asymptotic form of $\Sigma (i \omega_n)\rightarrow \Sigma_\infty + {\Sigma_0}/{i \omega_n}$ as $\omega_n \rightarrow \infty$.
As shown in Fig.~\ref{fig:1},
the causal self-energy obtained by our causal optimization is
identical to the exact self-energy at low frequencies, 
and it recovers the exact self-energy in the 
intermediate-frequency range which is not supplied as the input data,
indicating that the intermediate-frequency behavior
is not determined by system-dependent properties but by the general principle, causality.
This is consistent with previous work on
the sparsity of the imaginary-time data ~\cite{Shinaoka2017}
and allows us to obtain the self-energy practically in the entire 
frequency range using direct calculation at only low frequencies only and high-frequency asymptotic coefficients.

\begin{figure} 
\epsfxsize=8.5cm
\centering
\centerline{\epsfbox{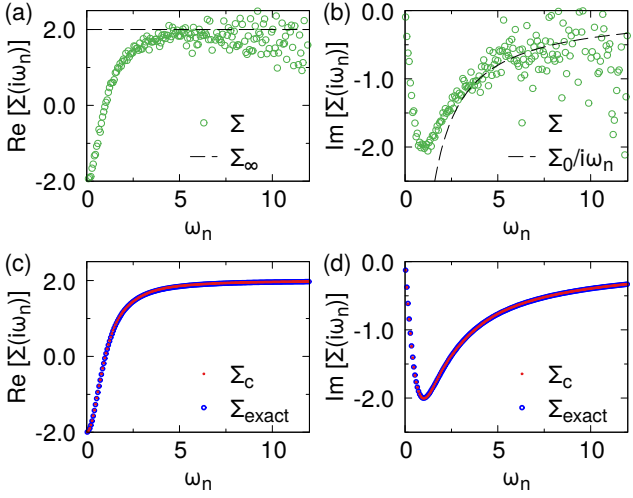}}
\caption {Causal optimization applied to the self-energy with statistical errors. (a) Real and (b) imaginary parts
of the statistically calculated noncausal self-energy $\Sigma$ of the Hubbard atom
which is the exact self-energy plus statistical errors. Dashed lines represent the
high-frequency asymptote.
(c) Real and (d) imaginary parts of the self-energy $\Sigma_{\textrm{c}}$
from our causal optimization method using $\Sigma$ and the high-frequency asymptote plotted in (a) and (b).
$\Sigma_{\textrm{c}}$ matches excellently with the exact self-energy $\Sigma_{\textrm{exact}}$.}
\label{fig:2}
\end{figure}

Many of many-body calculations rely on the QMC
method~\cite{Werner2006:1,Gull2011}. Due to its statistical nature, 
statistical errors are added to physical quantities.
Since our causal optimization eliminates noncausality
of the data,
we expect that our method can 
filter out statistical errors in QMC sampled data. To examine 
this idea, we consider a QMC procedure which converges the Green's function
to the exact one
$G_{\textrm{exact}}$ of Eq.~(\ref{eq:hat}) with $k_B T=0.01$, $n=0.5$, $\epsilon = -1$, and $U=4$.
To represent a practical QMC calculation, we sample the Green's function with Gaussian 
error~\cite{BoxMuller1958,Fernandez1996} so that, at each QMC step,
$G(i\omega_n) = G_{\textrm{exact}}(i\omega_n) + \Delta G_1 + i \Delta G_2$,
where $\Delta G_1,\Delta G_2 \sim \mathcal{N}(0,0.1^2)$, and it is averaged over $400$ sampling.
Then, the self-energy
$\Sigma (i \omega_n) = i \omega_n -\epsilon - G^{-1}(i \omega_n)$
has statistical errors increasing at high 
frequency [Figs.~\ref{fig:2}(a) and \ref{fig:2}(b)]. 
We applied our causal optimization on this self-energy with statistical errors
using the first $200$ frequency data and its high-frequency asymptotic coefficients as an input, obtaining
a smooth self-energy that coincides with the exact self-energy 
[Figs.~\ref{fig:2}(c) and \ref{fig:2}(d)].
Thus, causality filters out statistical errors.
There have been, so far, many approaches 
to filter such statistical errors in QMC
methods~\cite{Hafermann2012,Shinaoka2017,Boehnke2011,gull2018}.
Our causal optimization method can be used together with these methods, 
improving error-filtering effect.

\begin{figure} 
\epsfxsize=8.5cm
\centering
\centerline{\epsfbox{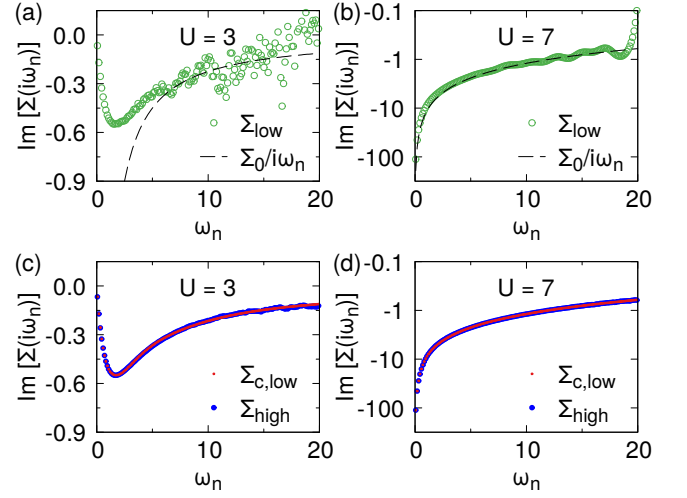}}
\caption {Causal optimization applied to the self-energy from DMFT.
[(a) and (b)] Imaginary parts of the self-energy $\Sigma_{\textrm{low}}$ from low-precision DMFT
calculation of the half-filled single-orbital
Hubbard model with (a)~$U=3$ and (b)~$U=7$. Dashed lines represent
the high-frequency asymptote from low-precision DMFT.
[(c) and (d)] Imaginary parts of the self-energy $\Sigma_{\textrm{c,low}}$ from our causal optimization 
method using only fifty low-frequency values of $\Sigma_{\textrm{low}}$
and the high-frequency asymptote shown in (a) and (b).
In (c) and (d), $\Sigma_{\textrm{c,low}}$ matches excellently with the high-precision self-energy 
$\Sigma_{\textrm{high}}$.}
\label{fig:3}
\end{figure}

\subsection{Single-orbital Hubbard model}

Using the finding that causality determines the self-energy 
at intermediate frequency and filters out statistical errors, we can improve efficiency of imaginary-time simulations
as follows.
First, we directly compute the low-frequency self-energy $\Sigma(i\omega_n)$ and
high-frequency asymptotic coefficients $\Sigma_{\infty}$ and $\Sigma_0$.
Then, we obtain the self-energy in other frequencies with our causal optimization method.
Because it is generally difficult to obtain high precision in the intermediate frequencies,
our approach can greatly reduce computing time compared with direct computation 
in wide range of $\omega_n$.
To show this, we conducted a DMFT simulation of the half-filled single-orbital Hubbard model~\cite{Hubbard1963}
described by Hamiltonian
\begin{equation}
H = -\!\!\sum_{<ij>\sigma} \!\! t_{ij} d^{\dagger}_{i\sigma} d^{ }_{i\sigma} 
   \!\!+\!\!  \sum_{i} U n_{i\uparrow} n_{i\downarrow}
   \!\!-\!\! \sum_{i} \mu (n_{i\uparrow} + n_{i\downarrow}).
\end{equation}
Here $d^{\dagger}_{i\sigma}$ ($d^{ }_{i\sigma}$) is the creation 
(annihilation) of an electron with spin $\sigma$ at site $i$, 
$n_{i\sigma}=d^{\dagger}_{i\sigma}d_{i\sigma}$,
$U$ is the Coulomb interaction strength,
$t_{ij}$ is electronic hopping parameters, 
and $\mu$ is the chemical potential.
We consider the infinite-dimensional Bethe lattice that has a
semicircular noninteracting density of states~\cite{Georges1996}.
This model has been extensively studied for the Mott transition~\cite{Georges1996}.
In our energy unit, noninteracting bandwidth is 4.
As an impurity solver, we implemented the hybridization expansion continuous-time QMC method~\cite{Werner2006:1}.
At temperature of $0.02$, we considered a paramagnetic metallic ($U=3$) phase
and a paramagnetic insulating ($U=7$) phase.
For each phase, we performed two calculations. One is a {\em low-precision} 
calculation, where only $6.4 \times 10^{7}$ QMC sampling is used for 
each iteration, followed by our causal optimization of the self-energy using
the low-precision results at the first $50$ frequencies and high-frequency asymptotic coefficients.
The other is a {\em high-precision} calculation where
the QMC sampling is $3600$ times the sampling in the low-precision calculation.
With the high precision, $\Sigma(i\omega_n)$ is obtained 
at the first $200$ frequencies. As shown clearly in Figs.~\ref{fig:3}(c) and \ref{fig:3}(d), 
results from our causal optimization of low-precision data
almost coincide with high-precision data, 
for both metallic and insulating phases.

\begin{figure} 
\epsfxsize=8.5cm
\centering
\centerline{\epsfbox{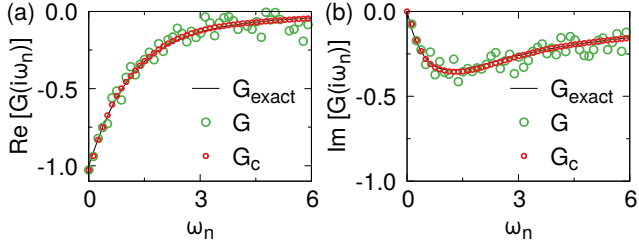}}
\caption {Causal optimization applied to the bosonic Green's function.
(a) Real and (b) imaginary parts of a bosonic Green's function $G$ with statistical errors
and its causal optimization $G_{\textrm{c}}$, shown in green and red dots,
respectively.
The Green's function $G$ is generated by adding Gaussian errors with the standard deviation of $0.05$
to the exact Green's function $G_{\textrm{exact}}$ shown in the black line. 
The exact Green's function $G_{\textrm{exact}}$ is 
generated by Eq.~(\ref{eq:bpd}) using $B(x)$ equal to the Gaussian distribution function with the mean of $1$
and the standard deviation of $0.8$.}
\label{fig:4}
\end{figure}

\section{\label{sec:II}Causal optimization for bosonic functions}

In this section, we derive the causal optimization method for the bosonic Green's function.
The causal bosonic Green's function $G(i\omega_n)$ satisfies
\begin{equation}
   G(i\omega_n) = \int  \frac{B(x) x}{i\omega_n -x} dx,\quad  B(x) \geq 0,  \label{eq:bpd}
\end{equation}
where $\omega_n = 2n \pi k_B T$ is the bosonic Matsubara frequency.
In the imaginary time, it is represented as
\begin{equation}
    G(\tau) = -\int  {B(x) x n_B(x) e^{(\beta-\tau)x}} d x , \quad B(x) \geq 0, \label{eq:btau}
\end{equation}
where $0\leq\tau\leq\beta$ and $n_B$ is
the Bose-Einstein distribution function $1/(e^{\beta x}-1)$~\cite{Abrikosov}.
Then, $x n_B(x) = x/(e^{\beta x}-1)$ and it is always positive.
As a result, Eq.~(\ref{eq:btau}) gives
\begin{equation}
     G^{(2k)}(\tau) \leq 0  \text{ for } k = 0, 1, 2, \cdots. \label{eq:bct}
\end{equation}
Thus, the causal optimization implemented for the fermionic case can be applied to the
bosonic Green's function, too.
In addition, since the correlation function $\chi(i\omega_n)$ satisfies
\begin{equation}
   \chi(i\omega_n) = -\int  \frac{B(x) x}{i\omega_n -x} dx,\quad  B(x) \geq 0,  \label{eq:cpd}
\end{equation}
which is the same with Eq.~(\ref{eq:bpd}) except for the minus sign in front of the integral, 
our causal optimization method can also be applied to $-\chi$.
Figure~\ref{fig:4} shows an example of the causal optimization of the bosonic Green's function.
In this example, we performed the causal optimization of a bosonic Green's function with statistical errors at $k_BT = 0.02$
using the first $100$ frequency data and the high-frequency asymptote $G(i\omega_n)\rightarrow 1/i\omega_n$.

\begin{figure} 
\epsfxsize=8.5cm
\centering
\centerline{\epsfbox{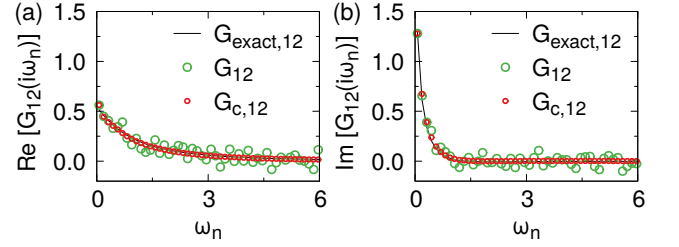}}
\caption {Causal optimization applied to a matrix-valued Green's function.
(a) Real and (b) imaginary parts of an off-diagonal element $G_{12}$ 
of a $2\times 2$ matrix-valued Green's function with statistical errors
and its causal optimization $G_\textrm{c,12}$, shown in green and red dots,
respectively.
The off-diagonal element $G_{12}$ is generated by adding Gaussian errors 
with the standard deviation of $0.05$ to the exact off-diagonal element 
$G_\textrm{exact,12}$ shown in the black line.
The exact $2\times 2$ matrix-valued Green's function 
$\mathbf{G}_\textrm{exact}$ is obtained by
$\mathbf{G}_\textrm{exact} = \mathbf{U}^\dagger \mathbf{G}^0\mathbf{U}$ with
$\mathbf{U} = e^{-i\frac{\pi}{4}\bm{\sigma}_2}$, $\bm{\sigma}_2=\big(
\begin{smallmatrix}
  0 & -i\\ 
  i & \hfill 0
\end{smallmatrix}\big)$,
and a diagonal matrix $\mathbf{G}^0$ of 
$G^0_{11}(i\omega_n) = 0.5 F(0,0.1,i\omega_n) + 0.5 F(2,0.3,i\omega_n)$ and $G^0_{22}(i\omega_n) = F(-1,0.5,i\omega_n)$.
Here $F(\mu,s,i\omega_n)$ is the Green's function with the Gaussian spectral function of the mean $\mu$ 
and the standard deviation $s$.}
\label{fig:5}
\end{figure}

\section{\label{sec:III}Causal optimization for matrix-valued functions}

Our causal optimization method can be extended
to matrix-valued Green's functions.
For a matrix-valued Green's function $\mathbf{G}(\tau)$, 
Eq.~(\ref{eq:ct}) is generalized to the definiteness condition
\begin{equation}
     \mathbf{G}^{(2k)}(\tau) \preceq 0  \text{ for } k = 0, 1, 2, \cdots. \label{eq:mct}
\end{equation}
Here $\preceq 0$ means negative semidefinite.
When $G_{ij}(i\omega_n)$ is the $(i,j)$ element of a 
numerically obtained matrix-valued Green's function,
we obtain the causal optimization $G_{\textrm{c},ij}(i\omega_n)$
of $G_{ij}(i\omega_n)$ by minimizing the distance 
$d_{ij}=(k_B T)^2 \sum_{i \omega_n} 
|G_{ij}(i\omega_n)-G_{\textrm{c},ij}(i\omega_n)|^2$.
Detailed description of our causal optimization method for 
matrix-valued functions is presented in Appendix B.

As an example, we consider a two-orbital Green's function $\mathbf{G}(i\omega_n)$.
First, we find the causal optimization of diagonal elements $G_{11}(\tau)$ and $G_{22}(\tau)$.
Then, the matrix $\mathbf{G}^{(2k)}(\tau)$ is negative definite if and only if
\begin{equation}
  G^{(2k)}_{22}(\tau)-G^{(2k)}_{12}(\tau) G^{(2k)}_{21}(\tau)/G^{(2k)}_{11}(\tau) < 0. \label{eq:2ct}
\end{equation}
We find the causal optimization $G_{\textrm{c},12}(i\omega_n)$ of $G_{12}(i\omega_n)$
by minimizing $d_{12} \!=\! (k_B T)^2 \sum_{i \omega_n}\!\!
|G_{12}(i\omega_n)-G_{\textrm{c},12}(i\omega_n)|^2$
under the quadratic constraint Eq.~(\ref{eq:2ct}).
Figure~\ref{fig:5} shows an example of the causal optimization of $G_{12}(i\omega_n)$.
In this example, we calculated the causal optimization of a matrix-valued Green's function
with statistical errors at $k_BT = 0.02$
using the first $100$ frequency data and the high-frequency asymptote $G_{ij}(i\omega_n)\rightarrow \delta_{ij}/i\omega_n$.

A single-particle fermionic matrix-valued Green's function can be represented as
$\mathbf{G}(i\omega_n) = 
(i \omega_n - \mathbf{t} - \mathbf{S}(i\omega_n))^{-1}$
for a causal function $\mathbf{S}(i\omega_n)$~\cite{Potthoff2007}.
Here $\mathbf{t}$ is a Hermitian matrix.
Then, in principle, the causality of $\mathbf{G}(i\omega_n)$
is equivalent with the causality of $\mathbf{S}(i\omega_n)$~\cite{Potthoff2007}.
In our numerical method, Eq.~(\ref{eq:ct}) is less restrictive than Eq.~(\ref{eq:tau})
so that $\mathbf{G}(i\omega_n)$ satisfying Eq.~(\ref{eq:ct})
does not guarantee $\mathbf{S}(i\omega_n)$ to satisfy Eq.~(\ref{eq:ct}).
On the other hand, $\mathbf{S}(i\omega_n)$ satisfying Eq.~(\ref{eq:ct})
always produces $\mathbf{G}(i\omega_n)$ satisfying Eq.~(\ref{eq:ct}) in every case we tested.
Thus, our causal optimization method performs better when
applied to $\Delta(z)$ or $\Sigma(z)$ than applied to $G(z)$.

\section{Stability of physical solution of the Luttinger-Ward functional}

\begin{figure} 
\epsfxsize=8.5cm
\centering
\centerline{\epsfbox{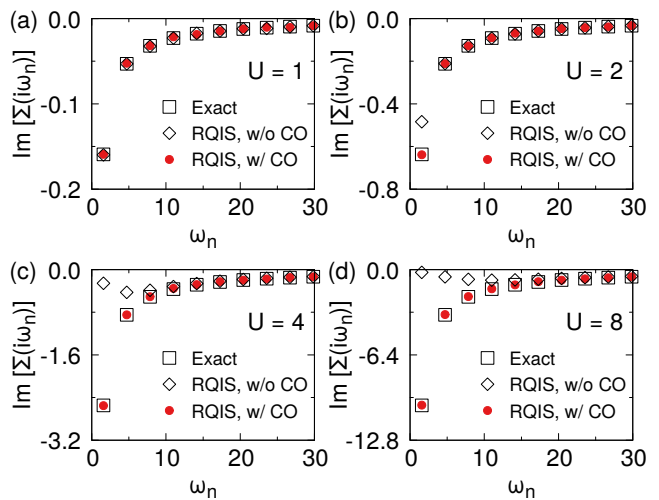}}
\caption{Causal optimization applied to stabilizing the physical solution of the Luttinger-Ward functional.
[(a)-(d)] Imaginary parts of the self-energy of the half-filled Hubbard atom for (a)~$U=1$, (b)~$U=2$, (c)~$U=4$, and
(d)~$U=8$. 
In (b)-(d), the self-energy calculated from the RQIS method without the causal optimization,
shown in empty diamonds, 
deviates from the exact self-energy, shown in empty squares, at low frequencies.
In contrast, the RQIS method with the causal optimization, shown in red dots, reproduces
the exact self-energy.}
\label{fig:6}
\end{figure}

Lastly, we consider the Luttinger-Ward functional~\cite{Luttinger1,Baym1961}.
We show below that our causal optimization method
can suppress the unphysical branch of the Luttinger-Ward functional.
First, to reproduce unphysical branches of the Luttinger-Ward functional
as in Ref.~\cite{Kozik2015}, we consider the Hubbard atom [Eq.~(\ref{eq:Hhat})]
with $k_B T=0.5$, $n=0.5$, $\epsilon = -U/2$, and various $U$ values,
and search for unphysical branches as follows.
From the exact Green's function $G$ [Eq.~(\ref{eq:hat})],
we calculate $G_0$, which can be physical or unphysical but gives 
$G$ correctly, using the reverse quantum impurity 
solver (RQIS) method~\cite{Vucicevic2018} where new estimate of $G_0$ is given by
\begin{equation}
    G^{\{i+1\}}_0(i\omega_n) = (G_{\textrm{exact}}^{-1}(i \omega_n) + \Sigma(i \omega_n)[G^{\{i\}}_0])^{-1}.\label{eq:g0update}
\end{equation}
For the \textit{i}th iteration, we use the interaction-expansion QMC
method~\cite{Rubtsov2005} to calculate the self-energy from $G^{\{i\}}_0$.
We start the iteration with the physical solution, $G^{\{1\}}_0 = 1/(i \omega_n-\epsilon)$.
If converged $G^{\{i\}}_0$ is different from this physical solution, then an unphysical 
branch is found for the given $G_{\textrm{exact}}$, together
with an unphysical self-energy $\Sigma(i \omega_n)$.
Figure~\ref{fig:6} compares $\Sigma(i \omega_n)$ calculated from
the RQIS method and
the physical self-energy ($\Sigma_{\textrm{exact}}(i \omega_n) = U/2 + U^2/4i\omega_n$).
The RQIS method produces the exact solution
for small interaction ($U=1$), but it produces unphysical 
solutions for larger interactions ($U=2,4,8$) as in Ref.~\cite{Kozik2015}, 
showing instability of the physical solution triggered by small statistical errors.

To examine the role of causality on the stability of physical solutions of the Luttinger-Ward functional, 
we enforced the causality of $G_0(i\omega_n)$ of Eq.~(\ref{eq:g0update}) 
by applying our causal optimization method to the hybridization function $\Delta(i\omega_n)$ for each iteration (see Appendix C for details).
Here $\Delta(i\omega_n)$ satisfies $G^{-1}_0(i\omega_n) = i\omega_n -\epsilon - \Delta(i\omega_n)$.
Then, calculational results with our causal 
optimization converge to the exact physical solution (Fig.~\ref{fig:6}), 
independently of starting values of $G_0^{\{1\}}(i\omega_n)$.
So, we can follow
the physical branch of the Luttinger-Ward functional by enforcing the causality of 
$G_0(i\omega_n)$ by applying our causal optimization method to $\Delta(i\omega_n)$.
In contrast, applying our causal optimization method directly to $G_0(i\omega_n)$
using Eq.~(\ref{eq:ct}) is not enough to make $G_0$ converge to the physical one in the RQIS method (see Appendix C for details),
which is related to the above discussion that Eq.~(\ref{eq:ct}) is less restrictive than Eq.~(\ref{eq:tau}).
We also note the causal optimization of the self-energy 
does not avoid unphysical solutions in the RQIS method and it is because unphysical 
self-energies in this problem are causal.
The set of physically accessible self-energies is different 
from the set of causal functions~\cite{Potthoff2012,Potthoff2003}.
By restricting the noninteracting Green's function to be causal, 
we constrain the self-energy to be physical as in Fig.~\ref{fig:6}.

\section{Summary}

In summary, we developed a causal optimization method
that ensures the causality of the imaginary-time Green's function and related quantities,
investigating practical consequences of the causality in imaginary-time simulations.
First, we verified that ensuring the causality extends the
low-frequency self-energy $\Sigma(i\omega_n)$ to the entire frequency range smoothly.
This property can be useful in quantum chemistry~\cite{Kananenka2016}
which requires numerical data in a wide range of imaginary frequencies.
Second, we showed the causality filters out statistical errors in QMC simulations.
Then, we used the causality to enhance computational
efficiency of practical QMC simulation.
This approach can be useful especially for the density functional theory
plus DMFT approach~\cite{Haule2010,Bhandary2016,Schler2018,Choi2019}.
Moreover, we demonstrated unphysical branches of the Luttinger-Ward functional
can be avoided by ensuring the causality of the noninteracting Green's function
using our causal optimization method.

\begin{acknowledgments}
This work was supported by NRF of Korea (Grant No. 2020R1A2C3013673), KISTI supercomputing center (Project No. KSC-2019-CRE-0195), and the Graduate School of Yonsei University Research Scholarship Grants in 2018.
\end{acknowledgments}

\appendix

\section{\label{sec:I}Numerical procedures of the causal optimization method}

In this section, we present the adaptive nonuniform grid, the cubic-spline interpolation,
and the quadratic programming used in our causal optimization method.

\subsection{Adaptive nonuniform grid}

\begin{figure} 
\epsfxsize=8.5cm
\centering
\centerline{\epsfbox{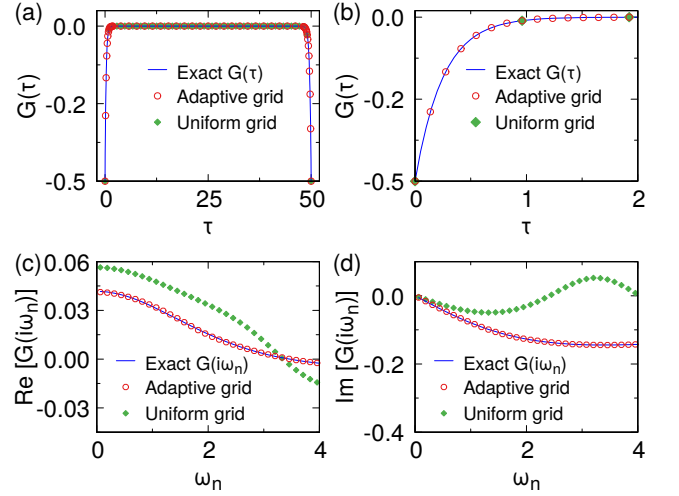}}
\caption {Comparison between adaptive- and uniform-grid representations of imaginary-time Green's function $G(\tau)$.
~(a) $G(\tau)$ in the imaginary time $\tau\in[0,\beta]$ with $\beta = 50$. Here
$G(\tau)$ corresponds to $G(i\omega_n) = \frac{1}{2} (\frac{1}{i\omega_n-4} + \frac{1}{i\omega_n+3})$.
(b) $G(\tau)$ of (a) plotted for $\tau\in[0,2]$.
(c) Real and (d) imaginary parts of imaginary-frequency Green's function $G(i\omega_n)$ obtained numerically
from $G(\tau_i)$ plotted in (a). In (c) and (d), $G(i\omega_n)$ obtained from the 
adaptive-grid representation of $G(\tau)$ matches with the exact $G(i\omega_n)$ excellently.
On the other hand, $G(i\omega_n)$ obtained from the uniform-grid representation of $G(\tau)$
fails to reproduce the exact $G(i\omega_n)$.}
\label{fig:7}
\end{figure}

We use nonuniform grid points $\{\tau_i\}$ to represent Green's function $G(\tau)$
in the imaginary time $\tau$.
Typically, an imaginary-time Green's function varies rapidly near $\tau=0$ and $\tau = \beta$.
Here $\beta=1/(k_BT)$.
An example of an imaginary-time fermionic Green's function is shown in Fig.~\ref{fig:7},
which corresponds to the imaginary-frequency Green's function
$G(i\omega_n) = \frac{1}{2} ( \frac{1}{i\omega_n-4} + \frac{1}{i\omega_n+3})$ at $k_B T = 0.02$.
Since many grid points are required in the range where $G(\tau)$ varies rapidly, 
we generate adaptive girds $\{\tau_i\}$ based on the equidistribution principle~\cite{Boor1},  
\begin{equation}
    \int_{\tau_i}^{\tau_{i+1}} |G^{(1)}(\tau)| \mathrm{d} \tau  = C/N_\tau, \, C = \int_{0}^\beta |G^{(1)}(\tau)| \mathrm{d} \tau. \label{eq:s1}
\end{equation}
Here $N_\tau$ is the number of grid points and $G^{(k)}$ is the $k$th derivative of $G$.
This method requires $|G^{(1)}(\tau)|$ after the causal optimization, and strict use of 
Eq.~(\ref{eq:s1}) may generate too sparse grid points except
for vicinity of $\tau=0$ and $\tau = \beta$.
Thus, we generate grid points as follows in practice.
First, we generate $n$ equidistant grid points $\{\tau^0_i\}$ in $[0,\beta]$ 
and calculate the causal optimization $G_{\textrm{c}}(\tau)$ using $\{\tau^0_i\}$.
Then we calculate $c_i = \int_{\tau^0_i}^{\tau^0_{i+1}} |G^{(1)}_{\textrm{c}}(\tau)| d\tau$ and
$m_i$ which is the integer nearest to $n c_i /\sum_{i=1}^{n-1} c_i$.
Finally, we divide each interval $[\tau^0_i,\tau^0_{i+1}]$ into $m_i+1$ intervals of the same width.
As a result, we have $N_\tau\approx 2 n$ grid points.
To describe the case with particle-hole symmetry correctly,
we generate grid points symmetric with respect to $\tau=\beta/2$.
With this adaptive grid, we perform the causal optimization again.
The advantage of our adaptive grid is represented in Fig.~\ref{fig:7}, 
where adaptive and uniform grids are generated with $N_\tau = 53$ and the adaptive grid describes the
imaginary-frequency Green's function much better than the uniform grid.

\subsection{Cubic-spline interpolation}

When values of Green's function are given at grid points $\{\tau_i\}$,
we interpolate Green's function $G(\tau)$ at $\tau_i\leq\tau\leq\tau_{i+1}$
using the cubic spline~\cite{Boor2,GullPhD,Georges1996,Bergeron2011,Boor1},
\begin{equation}
    G(\tau) = a_{i} + b_{i} (\tau-\tau_i) + c_{i} (\tau-\tau_i)^2 + d_{i} (\tau-\tau_i)^3,
\end{equation}
which requires $4 N_\tau-4$ parameters $a_i, b_i, c_i,$ and $d_i$.
Given values of $G(\tau_i)$, and continuity of $G(\tau_i)$ and its first and
second derivatives give $4 N_\tau - 6$ conditions for $a_i, b_i, c_i,$ and $d_i$.
Then two more conditions are needed. Although a popular choice is the so-called 
natural boundary condition~\cite{Georges1996,BlumerPhD} requiring $G^{(2)}(0) = G^{(2)}(\beta) = 0$,
this condition together with the causality condition gives $G^{(2)}(\tau) = 0$ for all $\tau$
because $G^{(2)}$ is concave as a result of the causality.
Thus, instead, we use two boundary conditions 
\begin{equation}
\begin{array}{ll}
{\displaystyle G^{(1)}(0)\pm G^{(1)}(\beta) = M_1,\quad}\\
{\displaystyle G^{(2)}(0)\pm G^{(2)}(\beta) = -M_2,\quad}
\end{array}
\end{equation}
as in Ref.~\cite{GullPhD}. Here two parameters $M_1$ and $M_2$ are high-frequency asymptotic coefficients
and the plus (minus) signs refer to the fermionic (bosonic) Green's function.
If $M_1$ and $M_2$ are not given as input data, then they are determined by minimizing
$d$ of Eq.~(\ref{eq:nchi2}).

\subsection{Quadratic programming}

In the implementation of our causal optimization,
we find $G_\textrm{c}$ which is closest to a given Green's function $G$ by minimizing
the distance $d$ of Eq.~(\ref{eq:nchi2}).
This minimizing procedure corresponds to a quadratic programming.
In general, a quadratic programming is to optimize $Q = \frac{1}{2} x^{T} G x + c^{T} x$ with respect to $x$,
with linear equality constraints $Ex-f=0$ and linear inequality constraints $Ax-b\ge0$. 
Here $x$ and $c$ are $n$-dimensional vectors, $G$ is an $n\times n$ matrix, 
$E$ is an $n_e\times n$ matrix, $f$ is an $n_e$-dimensional vector,
$A$ is an $m\times n$ matrix, $b$ is an $m$-dimensional vector, and 
$z\geq0$ for a vector $z$ represents that each component of $z$ is non-negative.
Thus, $n_e$ is the number of equality constraints and $m$ is the number of inequality constraints.
The optimal point $x$ satisfies following Karush-Kuhn-Tucker (KKT) conditions:
\begin{eqnarray}
	\nonumber
  Gx - A^{T}\lambda - E^{T}\nu + c = 0, \\
	\nonumber
  Ax-b-y = 0, \\ 
	\nonumber
  y_i \lambda_i = 0, \, y\geq0,\, \lambda \geq0, \\
  Ex - f = 0. \label{eq:kkt}
\end{eqnarray}
Here auxiliary variables $\lambda$ and $\nu$ are called as KKT multipliers.
The slack variable $y$ is introduced to transform inequality constraints $Ax-b\geq0$ to
simple nonnegativity condition $y\geq0$.
We use the interior point algorithm that finds a solution by traveling the interior of the feasible region.
Starting from a point $(x,y,\lambda,\nu)$ 
satisfying $y\geq 0$ and $\lambda\geq0$,
we find a step $(\Delta x, \Delta y, \Delta \lambda, \Delta \nu)$ satisfying
\begin{eqnarray}
	\nonumber
  G\Delta x - A^{T}\Delta\lambda - E^{T}\Delta\nu = -r_d,\\
	\nonumber
  A\Delta x - \Delta y=  -r_p, \\ 
	\nonumber
  \lambda_i \Delta y_i + y_i \Delta \lambda_i = -\lambda_i y_i  + \sigma \mu, \\
  E\Delta x = -r_e.
\end{eqnarray}
Here $\mu = y^{T}\lambda/m$, $r_d = G x - A^{T} \lambda + c$, $r_p = Ax-y-b$, and $r_e = Ex-f$.
The centering parameter $\sigma$ controls interiority and can be determined heuristically~\cite{Nocedal}.
Finally, we obtain a new point $(x,y,\lambda,\nu) + \alpha (\Delta x, \Delta y, \Delta \lambda, \Delta \nu)$
by choosing $\alpha$ that keeps the inequality $y\geq0$ and $\lambda\geq0$.
The interior-point method can also be used for a general nonlinear optimization problem~\cite{Nocedal}
although we explained the method for a quadratic programming.

\section{Numerical procedure of causal optimization for 
matrix-valued Green's functions}

As presented in the main text, the causality of the 
matrix-valued Green's function $\mathbf{G}(\tau)$ leads to
\begin{equation}
     \mathbf{G}^{(2k)}(\tau) \preceq 0  \text{ for } k = 0, 1, 2, \cdots. \label{eq:mct}
\end{equation}
Because negative definiteness of a matrix $\mathbf{A}$ is equivalent with
the positive definiteness of the matrix $-\mathbf{A}$,
it is enough to discuss the method for constraining positive definiteness.
Suppose we have an $n\times n$ Hermitian matrix $\mathbf{M}_n$ which is known to be positive definite.
Then, we consider a new $(n+1) \times (n+1)$ matrix
\begin{equation}
  \mathbf{M}_{n+1}=
\left[
\begin{array}{cc}
\mathbf{M}_n & \mathbf{v} \\
\mathbf{v}^{\dagger} & h
\end{array}
\right]
\end{equation}
which is Hermitian.
Here $\mathbf{v}$ is an $n$-dimensional vector and $h$ is a real number.
Since $\mathbf{M}_n$ is positive-definite, $\mathbf{M}_{n+1}$ is positive-definite if
and only if the Schur complement $h-\mathbf{v}^{\dagger}\mathbf{M}_n^{-1}\mathbf{v}>0$~\cite{Boyd},
which can be expressed as quadratic constraint for $x_{i} = \textrm{Re}[v_{i}]$ and $y_{i} = \textrm{Im}[v_{i}]$,
\begin{equation}
  a_i (x_i^2 + y_i^2) + 2 \textrm{Re}[b_i] x_i + 2 \textrm{Im}[b_i] y_i + c_i < 0. \label{eq:quadconst}
\end{equation}
Here $a_i = (M_n^{-1})_{ii}$, $b_i = \sum_{j\ne i} (M_n^{-1})_{ij} v_j$, 
and $c_i = \sum_{j\ne i,k\ne i} \allowbreak v^{*}_j (M_n^{-1})_{jk} v_k - h$.
Using these, we obtain the causal optimization of a matrix-valued Green's function as follows.
First, we conduct the causal optimization of each diagonal element $G_{ii}$ for $i = 1, \dots, n$
by minimizing $d_{ii}=(k_B T)^2 \sum_{i \omega_n} |G_{ii}(i\omega_n)-G_{\textrm{c},ii}(i\omega_n)|^2$
under the inequality constraint of Eq.~(\ref{eq:ct}).
Then, we perform the causal optimization of $G_{i,i+1}$ for $i = 1, \dots, n-1$ by
minimizing $d_{ij}=(k_B T)^2 \sum_{i \omega_n} |G_{ij}(i\omega_n)-G_{\textrm{c},ij}(i\omega_n)|^2$
for $j=i+1$ under
Eq.~(\ref{eq:quadconst}) applied to the $2\times2$ matrix $G_{jk}$ with $j,k=i,i+1$.
Then, we perform the causal optimization of $G_{i,i+2}$ for $i = 1, \dots, n-2$ by minimizing $d_{ij}$ for $j=i+2$
under Eq.~(\ref{eq:quadconst}) applied to the $3\times3$ matrix $G_{jk}$ with $j,k=i,i+1,i+2$.
Likewise, we perform the causal optimization of $G_{i,i+3}$ and so on until all off-diagonal
elements are optimized.
As a result, we obtain the Green's function $\mathbf{G}_{\textrm{c}}(\tau)$ that satisfies Eq.~(\ref{eq:mct}).

\begin{figure} 
\epsfxsize=8.5cm
\centering
\centerline{\epsfbox{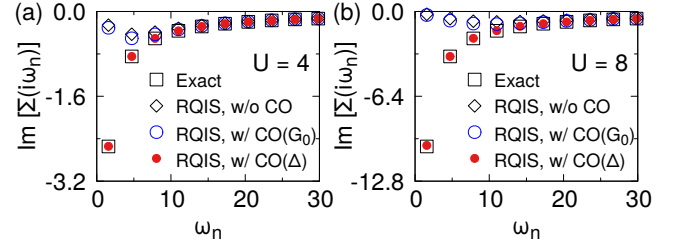}}
\caption {Causal optimization applied to stabilizing the physical solution of the Luttinger-Ward functional.
[(a) and (b)] Imaginary parts of the self-energy of the half-filled Hubbard atom for (a)~$U=4$ and (b)~$U=8$
at temperature of $0.5$. 
Empty squares represent the exact self-energy, and empty diamonds represent the self-energy calculated with
the RQIS method without the causal optimization.
Empty circles represent the self-energy from the RQIS with our causal optimization applied directly to $G_0$,
showing deviation from the exact self-energy. Red dots represent the self-energy from the RQIS with our causal optimization
applied to $\Delta$, showing excellent agreement with the exact self-energy.}
\label{fig:8}
\end{figure}

\section{\label{sec:IV}Comparison of causal optimization applied 
to $\Delta$ and applied directly to $G_0$}

In the RQIS applied to the half-filled Hubbard atom,
noninteracting Green's function $G_0$ is updated according to Eq.~(\ref{eq:g0update}).
Then, from $G^{-1}_0(i\omega_n)\allowbreak = i\omega_n - \epsilon - \Delta(i\omega_n)$,
the hybridization function $\Delta(i\omega_n)$ is updated as
\begin{equation}
    \Delta^{\{i+1\}}(i\omega_n) = \Sigma_{\textrm{exact}}(i \omega_n) - \Sigma^{\{i\}}(i \omega_n),\label{eq:deltaupdate}
\end{equation}
where $\Sigma^{\{i\}}(i\omega_n) = \Sigma(i\omega_n)[G^{\{i\}}_0]$.
From Eq.~(\ref{eq:deltaupdate}), the high-frequency asymptotic coefficient $M_0$ of $\Delta$, which satisfies
$\Delta(i \omega_n)\rightarrow{M_0}/{i \omega_n}$ 
as $\omega_n \rightarrow \infty$, is calculated as $M^{\{i+1\}}_0 = \Sigma_0^{\textrm{exact}} - \Sigma_0^{\{i\}}$,
where $\Sigma(i \omega_n) \rightarrow {\Sigma_0}/{i \omega_n}$ as $\omega_n \rightarrow \infty$.
Because $\Sigma_0$ depends on $n$ and $U$~\cite{Gull2011} only, 
$\Sigma_0^{\{i\}}$ is the same for every iteration and it is equal to $\Sigma_0^{\textrm{exact}}$,
resulting in $M_0^{\{i+1\}} = 0$ for every iteration, and
even the unphysical solution has the same high-frequency behavior with the physical one,
as shown in Fig.~\ref{fig:6}.
We note $M_0 = -(\Delta(0)+\Delta(\beta))$, so $\Delta(0)+\Delta(\beta)=0$.
With our causal optimization applied to $\Delta(i\omega_n)$, we have
$\Delta(\tau)\leq0$, so $\Delta(0)=\Delta(\beta)=0$.
Our causal optimization also restricts $\Delta(\tau)$ to be a concave function so that
$\Delta(\tau)\geq (1-\tau/\beta) \Delta(0) + (\tau/\beta) \Delta(\beta)=0$.
Thus, $\Delta(\tau)=0$ in the whole range of $0\leq \tau \leq \beta$
and $\Delta(i\omega_n)=0$ for all $\omega_n$,
which corresponds to the exact hybridization function of the Hubbard atom.
As a result, with our causal optimization, the RQIS converges to the exact solution
independent of the initial hybridization function $\Delta(i\omega_n)$.
In contrast, as mentioned in the main text, applying our causal optimization method directly to $G_0$ is not enough to make $G_0$
converge to the physical one in the RQIS method. Figure~\ref{fig:8} highlights difference in the RQIS results
with our causal optimization method applied directly to $G_0$ and that applied to $\Delta$.



%

\end{document}